\documentstyle[cite,aps,prl,epsfig,multicol]{revtex}

\def    \be             {\begin{equation}}
\def    \ee             {\end{equation}}
\def    \ba             {\begin{eqnarray}}
\def    \ea             {\end{eqnarray}}

\def    \=              {\;=\;}
\def    \frac           #1#2{{#1 \over #2}}

\newcommand\pt{p_{\rm \scriptscriptstyle T}}

\catcode`\@=11
\def\preprint#1{%
\def\@preprint{\noindent\flushright{#1}\vskip 10pt}%
}

\def \email#1{#1}

\begin{document}
\preprint{LPTHE-05-03\\
          Bicocca-FT-05-4\\
          LBNL-57063}


\title{QCD Predictions for Charm and Bottom Production at RHIC}

\author{Matteo Cacciari \\
LPTHE, Universit\'e Pierre et Marie Curie (Paris 6), France\\
E-mail: \email{cacciari@lpthe.jussieu.fr}
  }
  
\author{Paolo Nason \\
INFN, Sezione di Milano, Italy\\
E-mail: \email{Paolo.Nason@mib.infn.it}}
  
\author{Ramona Vogt\\
LBNL Berkeley and UC Davis, USA\\
E-mail: \email{vogt@lbl.gov}
  }

\maketitle

\begin{abstract}
We make up-to-date QCD predictions for open charm and bottom production
at RHIC in nucleon-nucleon collisions at $\sqrt{S} = 200$~GeV. 
We also calculate the electron spectrum resulting from heavy flavor decays
to allow direct comparison to the data. A rigorous benchmark,
including the theoretical uncertainties, is established against
which nuclear collision data can be compared to obtain evidence
for nuclear effects.
\end{abstract}

\begin{multicols}{2}



Over the past few years, heavy quark production at colliders
has received considerable attention since a variety of measurements
(photon-photon, photoproduction and
$p\bar p$ collisions) seemed to suggest a discrepancy, in particular 
for bottom production, with respect to
standard next-to-leading order (NLO) Quantum Chromodynamics (QCD)
predictions. New theoretical analyses and better measurements have,
however, largely reduced this discrepancy to the point that, in
most cases, it no longer appears significant
(see Refs.~\cite{Cacciari:2004ur,Mangano:2004xr} for a review and 
relevant references). 

It is important to continue to validate this
theoretical framework and its phenomenological inputs\footnote{We 
emphasize that such inputs were not chosen in order
to fit the data, but rather consistently extracted from other 
experimental measurements.} in new measurements such as the recent
heavy flavor data obtained at the Relativistic Heavy Ion
Collider (RHIC) by the PHENIX~\cite{Kelly:2004qw} and 
STAR~\cite{Adams:2004fc,Tai:2004bf} Collaborations.  Data taken
in $pp$ and d+Au collisions at $\sqrt{S_{NN}} = 200$ GeV and compared
to theoretical benchmark calculations will aid in the interpretation
of heavy flavor production in nucleus-nucleus collisions at the same
energies.  In these heavy ion collisions, which seek to establish the
existence of the quark-gluon plasma (QGP), a number of effects on heavy flavor
production have been predicted.  Of particular interest are effects which
modify the transverse momentum spectra of heavy flavor hadrons and their
decay products, including energy loss 
\cite{Djordjevic:2004nq,Armesto:2003jh,Dokshitzer:2001zm,Lin:1997cn}, 
transverse momentum broadening in both cold nuclear matter~\cite{Vogt:2001nh} 
and in passage through a hadronizing QGP~\cite{Svetitsky:1996nj} 
as well as collective effects such as transverse flow 
\cite{Greco:2003vf,Lin:2003jy}.  In addition, $J/\psi$ 
regeneration in a QGP from the initial
open charm yield has been suggested 
\cite{Thews:2000rj,Andronic:2003zv,Kostyuk:2003kt}.  Thus up-to-date
benchmark calculations of both the total charm yield and the transverse
momentum spectra are imperative.

The RHIC data are conveniently presented as real observables, either as
reconstructed hadronic decays of charm mesons or as the heavy flavor
decay electron spectra, with contributions from both charm and bottom
hadron decays.  In both cases, the transverse momentum spectra are
presented.  Such concrete observables, which can be directly compared to
predictions of the same quantities, stand in contrast to the often adopted
procedure of experimental `deconvolution' to a more basic level.
Such a deconvolution to the bare heavy quark level and the subsequent
extrapolation to full phase space, sometimes involving large factors,
risks biasing the data since theoretical prejudice enters in both the
deconvolution and the extrapolation, and should therefore be avoided.

The purpose of this paper is neither to review all possible methods
to evaluate the heavy quark cross section in heavy ion collisions, nor
to perform a detailed analysis of the data. Instead,
besides comparing the RHIC data to the most up-to-date QCD predictions,
we establish the aforementioned benchmark
calculation for further comparisons. 
To this end, we thus adhere to the rigorous QCD framework shown to be 
successful in $p\bar{p}$ collisions. Significant deviations from this 
benchmark could thus signal the presence of effects specific to
the high density environment of heavy ion collisions, such as those mentioned
previously. 

To make comparisons at various levels (while
preferring the final observable), in this letter we present predictions 
of the transverse momentum, $\pt$, distributions of
charm and bottom quarks, the charm and bottom hadron distributions 
resulting from fragmentation and,
finally, the electrons produced in semileptonic decays of the hadrons. 
At each step, we clarify the theoretical framework as well as the 
parameters and phenomenological inputs. Theoretical
uncertainties are estimated as extensively as possible since 
comparisons of data with theory should not be performed at the 
`central value' level only but should also include the respective 
uncertainties. Our final prediction is thus not a single curve but rather an
uncertainty band which has a reasonably large probability of containing
the `true' theoretical prediction\footnote{Since a rigorous statistical
estimate for this probability is not possible in the frequentist
sense for theoretical errors, one might resort to a Bayesian analysis
to quantify it. While refraining from such an effort, we
roughly estimate this probability as $\sim 80$-90\%.}.

The theoretical prediction of the electron spectrum includes 
three  main components: the $\pt$ and rapidity
distributions of the heavy quark $Q$ in $pp$ collisions at $\sqrt{S} =
200$~GeV, calculated in perturbative QCD; fragmentation of the
heavy quarks into heavy hadrons, $H_Q$, described by
phenomenological input extracted from $e^+e^-$ data; and the decay of
$H_Q$ into electrons according to spectra available from other 
measurements. This
cross section is schematically written as
\begin{eqnarray}
\frac{E d^3\sigma(e)}{dp^3} &=& \frac{E_Q d^3\sigma(Q)}{dp^3_Q} \nonumber \\
&\otimes& D(Q\to H_Q) \otimes f(H_Q \to e) \nonumber \; , 
\end{eqnarray}
where the symbol $\otimes$ denotes a generic convolution. The electron decay
spectrum term $f(H_Q \to e)$ also implicitly accounts for the proper 
branching ratio.

The distribution $E d^3\sigma(Q)/dp^3_Q$ is evaluated at the Fixed-Order plus
Next-to-Leading-Log (FONLL) level, implemented in Ref.~\cite{Cacciari:1998it}.
In addition to including the full fixed-order NLO
result~\cite{Nason:1987xz,Beenakker:1990ma}, the FONLL calculation also
resums~\cite{Cacciari:1993mq} large perturbative terms proportional to
$\alpha_s^n\log^k(\pt/m)$ to all orders with next-to-leading logarithmic (NLL)
accuracy (i.e. $k=n,\,n-1$) where $m$ is the heavy quark mass. The perturbative
parameters are the heavy quark mass and the value of the strong coupling,
$\alpha_s$. We take $m_c = 1.5$~GeV and $m_b = 4.75$~GeV as reference values
and vary the masses over the range $1.3< m_c <1.7$ GeV for charm and $4.5<m_b<
5$ GeV for bottom to estimate the resulting mass uncertainties. The QCD scale
at five flavors, $\Lambda^{(5)}$, is set to 0.226 GeV, i.e. 
the value provided by the
CTEQ6M parton densities. The perturbative calculation also depends on the
unphysical factorization ($\mu_F$) and renormalization ($\mu_R$) scales.  The
sensitivity of the cross section to their variation can be used to estimate the
perturbative uncertainty due to the absence of higher orders. We have taken
$\mu_{R,F} = \mu_0 = \sqrt{\pt^2 + m^2}$ as a central value and varied
the two scales independently within a  `fiducial' region defined by  $\mu_{R,F}
= \xi_{R,F}\mu_0$ with $0.5 \le \xi_{R,F} \le 2$ and $0.5 \le  \xi_R/\xi_F \le
2$. In practice, we use the following seven sets: $\{(\xi_R,\xi_F)\}$ =
\{(1,1),  (2,2), (0.5,0.5), (1,0.5), (2,1), (0.5,1), (1,2)\}.  The envelope
containing the resulting curves defines the uncertainty. Finally, the
uncertainties stemming from mass and scale variations are added in quadrature.

These `perturbative' inputs lead to a FONLL  total $c\bar c$ cross section in
$pp$ collisions of $\sigma_{c\bar c}^{\rm FONLL} = 256^{+400}_{-146}$~$\mu$b 
at $\sqrt{S} = 200$~GeV.  The theoretical uncertainty is evaluated as described
above. The corresponding NLO  prediction\footnote{Earlier
papers~\cite{Vogt:2001nh} used $m_c = 1.2$~GeV and $\mu_R = \mu_F =
2\sqrt{\pt^2 + m^2}$ as reference parameters. With this choice we find
$\sigma_{c\bar c}^{\rm NLO} = 427$~$\mu$b.}  is $244^{+381}_{-134}$~$\mu$b. 
Thus the two calculations are equivalent at the total cross section level 
within the large perturbative uncertainties, as expected.  The total cross
section for bottom production is $\sigma_{b\bar b}^{\rm FONLL} =
1.87^{+0.99}_{-0.67}$~$\mu$b.

The fragmentation functions $D(c\to D)$ and $D(b\to B)$, where $D$ and $B$
indicate a generic admixture of charm and bottom hadrons, are consistently
extracted from $e^+e^-$ data in the context of a FONLL-type calculation, as
described in Refs.~\cite{Cacciari:2002pa,Cacciari:2003zu,Cacciari:2003uh}. The
charm fragmentation function~\cite{Cacciari:2003zu} depends on the parameter
$r$~\cite{Braaten:1994bz} with $r = 0.1$ for $m_c = 1.5$~GeV, $r =
0.135$ for $m_c = 1.7$~GeV, and $r = 0.06$ for $m_c = 1.3$~GeV from $e^+e^-$
fits. Bottom fragmentation instead depends on the  parameter $\alpha$ in a
functional form by Kartvelishvili {\it et  al.}~\cite{Kartvelishvili:1977pi}:
$\alpha = 29.1$ for $m_b = 4.75$~GeV,  $\alpha = 34$ for $m_b = 5$~GeV, and
$\alpha = 25.6$ for $m_b = 4.5$~GeV (see Ref.~\cite{Cacciari:2003uh}). It is
worth noting that using the Peterson {\it et  al.} fragmentation
function~\cite{Peterson:1982ak}, with standard parameter choices $\epsilon_c
\simeq
0.06 \pm 0.03$ and $\epsilon_b \simeq 0.006 \pm 0.003$ would not provide 
a valid description of fragmentation in FONLL~\cite{Cacciari:2002pa}. 
Fragmentation is numerically performed by
rescaling the quark three-momentum at a constant angle in the laboratory frame.
This choice is, to some extent, arbitrary. Alternatively, one might rescale the
transverse momentum at constant rapidity. While all choices are equivalent at
$\pt \gg m$, they will, in general, lead to different results at $\pt \simeq
m$, where a large fraction of the RHIC data lie. The ensuing
uncertainty is, however, not larger than the perturbative
ones~\cite{Cacciari:2003uh} and will therefore not be considered in more
detail.

\begin{figure}
\begin{center}
\epsfig{file=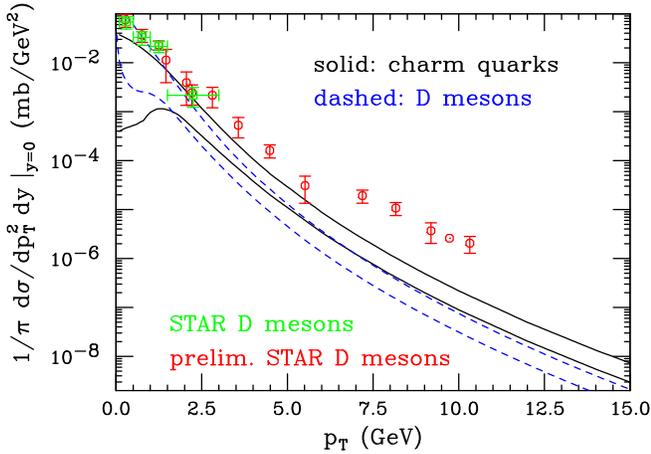,width=\columnwidth,clip=}
\caption{\small\label{cD} The theoretical uncertainty bands for the charm 
quark and $D$ meson $p_T$ distributions in $pp$ collisions 
at $\sqrt{S} = 200$~GeV, using BR($c \to D$) = 1. Data from the 
STAR Collaboration from d+Au collisions (scaled to $pp$ using $N_{\rm bin}$ =
7.5) at $\sqrt{S_{NN}} = 200$~GeV, 
final~\cite{Adams:2004fc} and preliminary~\protect\cite{Tai:2004bf},
are also shown.
}
\end{center}
\end{figure}

\begin{figure}
\begin{center}
\epsfig{file=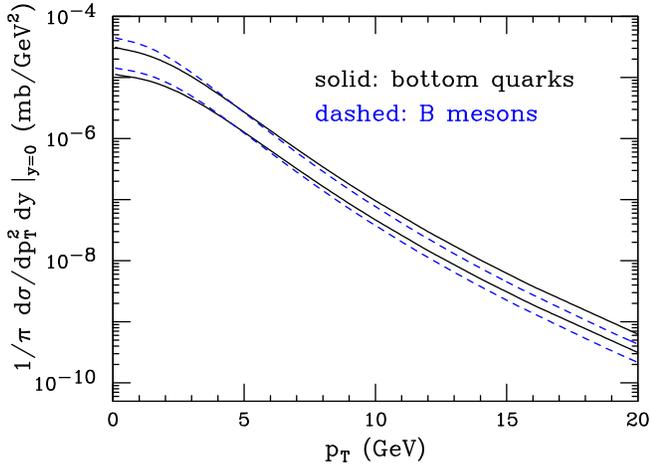,width=\columnwidth,clip=}
\caption{\small\label{bB} The theoretical uncertainty bands for the bottom 
quark and $B$ meson $p_T$ distributions in $pp$ collisions 
at $\sqrt{S} = 200$~GeV, using BR($b \to B$) = 1.}
\end{center}
\end{figure}

The decay of the $D$ and $B$ mesons into electrons is controlled by the
experimentally measured decay spectra and branching ratios (BR).
The spectrum for primary $B\to e$ decays has been measured recently by
BaBar\cite{Aubert:2004td} and CLEO\cite{Mahmood:2004kq}. We have used
a model that fits the data well and assume it to be valid for all
bottom hadrons.  Preliminary CLEO data on the inclusive electron 
spectrum in semi-leptonic $D$
decays have been shown~\cite{yelton}. We fitted this spectrum and we assume 
it to be identical for all charm hadrons. Finally, the contribution of 
electrons from the
secondary $B$ decays $B\to D\to e$ has also been accounted for.  The
relevant electron spectrum has been obtained
as a  convolution of the $D\to e$ spectrum mentioned above
with a parton-model prediction for the $b\to c$ decay.
The resulting electron spectrum is very soft,
suggesting that its contribution to the total
will most likely be negligible.

To normalize the decay spectra, we use the branching ratios
for bottom and charm hadron mixtures\footnote{Note that the experimental 
uncertainties on the BRs have
not been included in the overall uncertainty of the predictions, since they are
much smaller than those of perturbative origin.} appropriate to this
high energy regime~\cite{Eidelman:2004wy}:
BR$(B\to e) = 10.86 \pm 0.35$\%, BR$(D\to e) = 10.3 \pm 1.2$\%,
and BR$(B\to D\to e) = 9.6 \pm 0.6$\%.

We first present the transverse momentum
distributions for charm quarks and charm hadrons. Figure~\ref{cD} 
shows the theoretical uncertainty bands for the
two distributions, obtained by summing the mass and scale uncertainties
in quadrature\footnote{
For example, the value of the upper curve is
${\rm U}(\pt) = 
 {\rm C}(\pt)
+\sqrt{({\rm M}_s(\pt)-{\rm C}(\pt))^2 + ({\rm M}_m(\pt)-{\rm C}(\pt))^2 }$, 
where C stands for the central values,
${\rm M_s}$ is the maximum cross section
obtained by choosing $m_c=1.5$~GeV with the scale factors 
in our seven fiducial sets,
and ${\rm M_m}$ is the maximum cross section obtained
with $\xi_R=\xi_F=1$ and $m_c=1.3,\;1.5$ and $1.7$~GeV.}.
Note that the band is enlarged at low
$\pt$ due to the large value of $\alpha_s$ at low
scales and the increased sensitivity of the cross section to the
charm quark mass. It is also worth noting that, due to the fairly hard
fragmentation function,  the $D$ meson and $c$ quark
distributions begin to differ outside the uncertainty bands 
only for $\pt > 9$~GeV. 
The same comparison is shown in Fig.~\ref{bB} for bottom quarks and
the subsequent $B$ mesons. As a result of the harder $b\to B$ fragmentation
function, the two bands partially overlap for 
$\pt \simeq 20$~GeV and beyond.

\begin{figure}[t]
\begin{center}
\epsfig{file=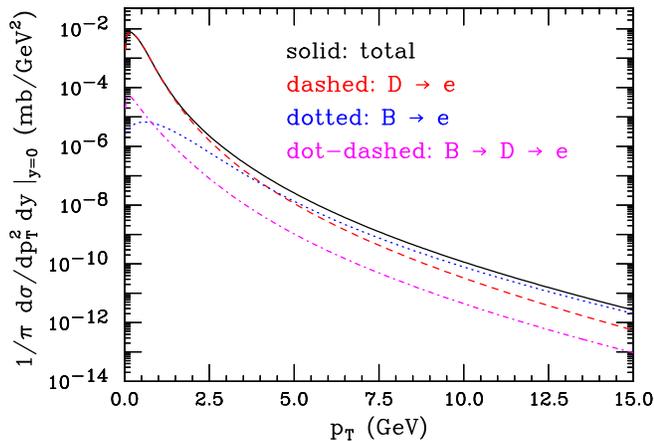,width=\columnwidth,clip=}
\caption{\small\label{components} The various components of the electron
transverse momentum spectrum. These curves are calculated with 
the central
masses and scales, i.e. $m_c = 1.5$~GeV, $m_b = 4.75$~GeV and $\xi_{R,F} = 1$.}
\end{center}
\end{figure}

We next consider the transverse momentum distributions of electrons
from $D$ and $B$ decays. Figure~\ref{components} shows the
contributions from $D\to e$, $B \to e$ and $B\to D \to e$ decays as well
as the total.  As anticipated, the
$B\to D\to e$ secondary electron spectrum is extremely soft, only
exceeding the primary $B\to e$ decays at $\pt<1$ GeV.  It is always 
negligible with respect to the total yield. We
further note that the electron spectrum from $B$ decays
becomes larger than that of electrons from $D$ decays at $\pt \simeq 4$~GeV.
The qualitative features of this plot are in good agreement 
with results obtained by the RHIC Collaborations using the 
PYTHIA event generator~\cite{Sjostrand:2000wi}.

\begin{figure}[t]
\begin{center}
\epsfig{file=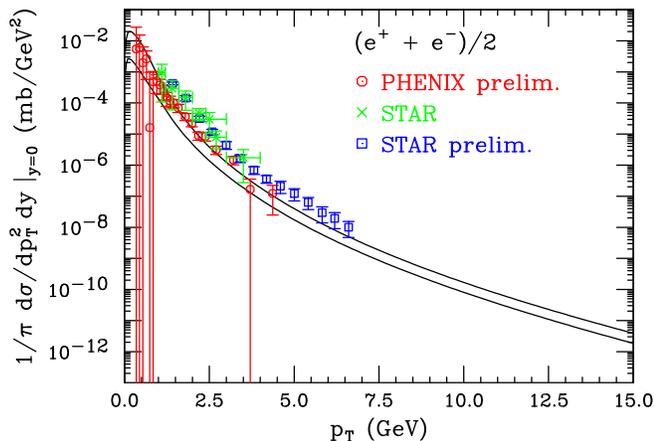,width=\columnwidth,clip=}
\caption{\small\label{electrons} The final prediction for the
theoretical uncertainty band of the electron spectrum from charm and
bottom in $pp$ collisions. 
Preliminary data from the PHENIX~\protect\cite{Kelly:2004qw} 
and STAR (final~\protect\cite{Adams:2004fc} and preliminary~\protect\cite{Tai:2004bf})
Collaborations are also shown.}
\end{center}
\end{figure}

Finally, in Fig.~\ref{electrons} we show the theoretical uncertainty
band for electrons coming from charm and bottom hadron decays at
$\sqrt{S} = 200$~GeV at RHIC. The sum of the three components shown in 
Fig.~\ref{components} corresponds to the central value of the band in 
Fig.~\ref{electrons}. The upper and lower limit of the 
band are obtained by summing the upper and lower limit for each component.

In conclusion, we have evaluated the higher order QCD charm and bottom quark
production cross sections in $\sqrt{S} = 200$~GeV $pp$ collisions at RHIC.  The
results are presented in the form of a theoretical uncertainty band for the
transverse momentum distribution of  either bare charm (bottom), $D$ ($B$)
mesons, or electrons originating from the decay of charm and bottom hadrons.
This result should not be multiplied by any $K$ factor before comparison with
data. Rather, agreement within the uncertainties of the measurements will
support the applicability of standard QCD calculations to heavy quark
production at RHIC. Alternatively, a significant disagreement will suggest the
need to complement this evaluation with further ingredients.

\acknowledgments{We wish to thank Jack Smith for having suggested the
calculation of these predictions, John Yelton and David Asner for
information on the CLEO-c preliminary data for inclusive $D\to e$ decay,
Ralf Averbeck, Sergey Butsyk, Gouranga Nayak and An Tai for many details 
about the ongoing PHENIX and STAR heavy flavour analyses.}

\end{multicols}
\end{document}